\documentclass[journal]{IEEEtran}

\usepackage{graphicx} \usepackage{bm} \usepackage{amsmath}
\usepackage{cite}
\usepackage{amssymb} \newcommand{\etal}{{\it et al. }}

\usepackage{amsmath,amsthm,amssymb,mathrsfs}
\usepackage{bm}
\usepackage{subfigure}

\newcommand{\Plog}{{\rm plog}}
\newcommand{\D}{{\rm d}}
\newcommand{\E}{{\rm e}}

\ifCLASSINFOpdf
\else
\fi
\begin{document}

\title{Theory of Spin Torque Assisted Thermal Switching of Single Free Layer}
\author{Tomohiro~Taniguchi
        and~Hiroshi~Imamura
        \\
        Nanosystem Research Institute (NRI), 
        National Institute of Advanced Industrial Science and Technology (AIST),
        \\
        Tsukuba, Ibaraki 305-8568, Japan
\thanks{${}^{*}$Corresponding author. Email address: h-imamura@aist.go.jp}}

\maketitle

\begin{abstract}
%\boldmath
The spin torque assisted thermal switching of the single free layer was studied theoretically. 
Based on the rate equation, 
we derived the theoretical formulas of 
the most likely and mean switching currents 
of the sweep current assisted magnetization switching, 
and found that the value of the exponent $b$ in the switching rate formula  significantly affects 
the estimation of the retention time of magnetic random access memory. 
Based on the Fokker-Planck approach, 
we also showed that the value of $b$ should be two, 
not unity as argued in the previous works. 
\end{abstract}


\IEEEpeerreviewmaketitle

% =========================================================================================== %

\section{Introduction}
\label{sec:Introduction}

\IEEEPARstart{M}{agnetic} random access memory (MRAM) 
using tunneling magnetoresistance (TMR) effect [1],[2] and spin torque switching [3],[4] has attracted much attention
for spintronics device applications 
due to its non-volatility and fast writing time with a low switching current. 
A high thermal stability ($\Delta_{0}$) (more than 60) of magnetic tunnel junctions (MTJs)
is also important to keep the information in MRAM more than ten years. 
Recently, Hayakawa \etal [5] and Yakata \etal [6],[7] respectively reported that 
the anti-ferromagnetically (AF) and ferromagnetically (F) coupled 
synthetic free layers show high thermal stabilities 
($\Delta_{0}>80$ for AF coupled layer and $\Delta_{0}=146$ for F coupled layer) 
compared to a single free layer. 

% =========================================================================================== %

The thermal stability has been determined 
by measuring the spin torque assisted thermal switching of the free layer 
and analyzing the time evolution of the switching probability 
by Brown's formula [8] with the spin torque term. 
The theoretical formula of the switching probability is generally given by 
$P=1-\exp[-\int_{0}^{t} \D t^{\prime}\nu(t^{\prime})]$, 
where $\nu(t)=f_{0}\exp[-\Delta_{0}(1-I/I_{\rm c})^{b}]$. 
Here, $f_{0}$, $I$, and $I_{\rm c}$ are the attempt frequency, 
current magnitude, and critical current of the spin torque switching at zero temperature, respectively. 
$b$ is the exponent of the current term in the switching rate $\nu$, 
and was argued to be unity by Koch \etal in 2004 [9]. 
On the other hand, recently, Suzuki \etal [10] and we [11],[12] independently studied the spin torque assisted thermal switching theoretically, 
and showed that the exponent $b$ should be two. 
Since the estimation of the thermal stability strongly depends on the value of $b$, 
as discussed in this paper, 
the determination of $b$ is important for the spintronics applications. 

% =========================================================================================== %

In this paper, 
we study the spin torque assisted thermal switching of the single free layer theoretically. 
In Sec. \ref{sec:Theory of Magnetization Switching due to Sweep Current}, 
we derive the theoretical formulas of the most likely and mean switching currents 
of the sweep current assisted magnetization switching, 
and study the effect of the value of the exponent $b$ 
on the estimation of the retention time of the MRAM. 
In Sec. \ref{sec:Comparison with Theory of Koch etal}, 
the differences of the theories in Refs. [9],[10],[11] are discussed 
by analyzing the solution of the Fokker-Planck equation. 
Section \ref{sec:Conclusion} is devoted to the conclusions. 

% =========================================================================================== %

% =========================================================================================== %

% =========================================================================================== %

\section{Theory of Magnetization Switching due to Sweep Current}
\label{sec:Theory of Magnetization Switching due to Sweep Current}

In this section, we consider the spin torque assisted thermal switching 
of the uniaxially anisotropic free layer, 
which has two minima of its magnetic energy. 
At the initial time $t=0$, 
the system stays one minimum. 
From $t=0$, the electric current $I(t)=\varkappa t$ is applied to the free layer 
which exerts the spin torque on the magnetization and assists its switching. 
In this section, the current is assumed to increase linearly in time with the sweep rate $\varkappa$, 
as done in the experiments [7],[13],[14]. 
The magnitude of the current $I(t)=\varkappa t$ should be less than $I_{\rm c}$ 
because we are interested in the thermally activated region. 
The time evolution of the survival probability of the initial state, $R(t)$, is described by the rate equation, 
\begin{equation}
  \frac{\D R(t)}{\D t}
  =
  -\nu(t)
  R(t),
  \label{eq:rate_equation}
\end{equation}
where the switching rate $\nu(t)$ is given by 
\begin{equation}
  \nu(t)
  =
  f_{0}
  \exp
  \left[
    -\Delta_{0}
    \left(
      1
      -
      \frac{I(t)}{I_{\rm c}}
    \right)^{b}
  \right].
  \label{eq:switching_rate}
\end{equation}
We assume that the attempt frequency is constant. 
$b$ is the exponent of the current term, $(1-I/I_{\rm c})$. 
The switching probability is given by $P(t)=1-R(t)$. 
Also, we define the probability density $p(t)$ by 
$p(t)=-\D R/\D t=\D P/\D t$. 
Equation (\ref{eq:rate_equation}) describes the escape from one equilibrium to the others 
in many physical systems, 
and the value of $b$ reflects their energy landscape: 
$b=1$ for the Bell's approximation [15], 
$b=3/2$ for the linear-cubic potential [16],
and $b=2$ for the parabolic potential [17],[18]. 
The determination of the value of $b$ has been discussed
not only in spintronics but also the other fields of physics [19]. 
The form of Eq. (\ref{eq:switching_rate}) is the special case of the model of Garg 
($a$ in Ref. [20] corresponds to $1-b$). 

% =========================================================================================== %
% =========================================================================================== %

\begin{figure}
  \centerline{\includegraphics[width=1.0\columnwidth]{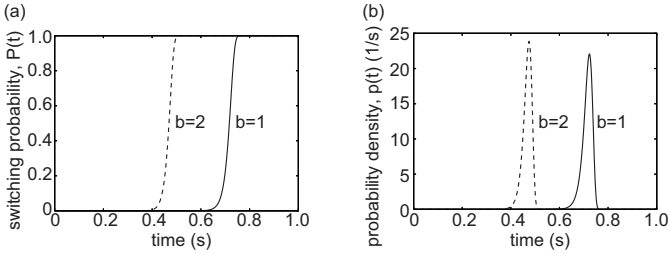}}
  \caption{
           The time evolutions of 
           (a) the switching probability $P(t)$ and (b) its density $p(t)$
           for $b=1$ (solid) and $b=2$ (dotted). 
  \vspace{-3.5ex}}
  \label{fig:fig1}
\end{figure}

% =========================================================================================== %
% =========================================================================================== %

% =========================================================================================== %

The solution of Eq. (\ref{eq:rate_equation}) with the initial condition $R(0)=1$ is given by 
\begin{equation}
  R(t)
  \!=\!
  \exp
  \left\{\!
    -\frac{f_{0}I_{\rm c}}{b \varkappa \Delta_{0}^{1/b}}
    \!\!
    \left[
      \gamma\!
      \left(\!
        \frac{1}{b},\Delta_{0}
      \!\right)
      \!-\!
      \gamma\!
      \left(\!
        \frac{1}{b}, \Delta_{0}\! \left(\! 1 \!-\! \frac{I}{I_{\rm c}} \!\right)^{b}
      \right)\!
    \right]\!
  \right\},
\end{equation}
where $\gamma(\beta,z)=\int_{0}^{z} \D t t^{\beta-1}\E^{-t}$ is the lower incomplete $\Gamma$ function. 
Figure \ref{fig:fig1} the time evolutions of 
(a) the switching probability $P(t)$ and (b) its density $p(t)$. 
The values of the parameters are taken to be 
$f_{0}=1.0$ GHz, $I_{\rm c}=1.0$ mA, $\varkappa=1.0$ mA/s, and $\Delta_{0}=60$, respectively, 
which are typical values found in the experiments [6],[7],[13],[14]. 
As shown, $P(t)$ suddenly changes from 0 to 1 at a certain time $t=\tilde{t}$ 
at which $p(t)$ takes its maximum. 
We call $\tilde{t}$ the switching time. 
The switching time $\tilde{t}$ is determined by the condition $(\D p(t)/\D t)_{t=\tilde{t}}=0$, 
i.e., $\D\nu/\D t=\nu^{2}$, 
and is given by 
\begin{equation}
  \frac{\varkappa \tilde{t}}{I_{\rm c}}
  =
  1
  -
  \frac{1}{\Delta_{0}}
  \log
  \left(
    \frac{f_{0}I_{\rm c}}{\varkappa \Delta_{0}}
  \right),
\end{equation}
for $b=1$,
and 
\begin{equation}
  \frac{\varkappa \tilde{t}}{I_{\rm c}}
  =
  1
  -
  \left\{
    \frac{b-1}{b\Delta_{0}}
    \Plog
    \left[
      \frac{b}{b-1}
      \left(
        \frac{f_{0}I_{\rm c}}{b \varkappa \Delta_{0}^{1/b}}
      \right)^{b/(b-1)}
    \right]
  \right\}^{1/b},
  \label{eq:most_likely_rupture_current_b_1}
\end{equation}
for $b>1$. 
Here $\Plog(z)$ is the product logarithm
which satisfies $\Plog(z)\exp[\Plog(z)]=z$. 
For a large $z \gg 1$, $\Plog(z) \simeq \log z$, 
and $\tilde{t}$ ($b>1$) can be approximated to 
\begin{equation}
  \frac{\varkappa \tilde{t}}{I_{\rm c}}
  \simeq 
  1
  -
  \left\{
    \frac{1}{\Delta_{0}}
    \log
    \left[
      \left(
        \frac{b}{b-1}
      \right)^{1-1/b}
      \frac{f_{0}I_{\rm c}}{b\varkappa \Delta_{0}^{1/b}}
    \right]
  \right\}^{1/b}.
  \label{eq:most_likely_rupture_current_b}
\end{equation}
The current at $t=\tilde{t}$, $I(\tilde{t})=\varkappa\tilde{t}$, is
the most likely switching current for the thermal switching. 
Since we are interested in the switching after the injection of the current at $t=0$, 
$\tilde{t}$ should be larger than zero. 
Thus, the above formula is valid in the sweep rate range $\varkappa > \varkappa_{\rm c}$, 
where the critical sweep rate $\varkappa_{\rm c}$ is given by 
\begin{equation}
  \varkappa_{\rm c}
  =
  \frac{f_{0}I_{\rm c}}{b\Delta_{0}}
  \E^{-\Delta_{0}}.
\end{equation}
The value of $\varkappa_{\rm c}$ estimated by using the above parameter values 
is on the order of $10^{-19}$ mA/s, 
which is much smaller than the experimental values ($0.01-1.0$ mA/s in Ref. [14]).
Thus, the above analysis is applicable to the conventional experiments. 

% =========================================================================================== %

We also define the mean switching current $\langle I \rangle$ by 
\begin{equation}
  \langle I \rangle 
  =
  \int_{0}^{1} \D R 
  I
  =
  -\int_{0}^{\infty} \D t 
  \frac{\D R}{\D t}
  \varkappa t
  =
  \varkappa \int_{0}^{\infty} \D t
  R. 
  \label{eq:average_current}
\end{equation}
Since $p(t)$ takes its maximum at $t=\tilde{t}$, 
we approximate that 
\begin{equation}
\begin{split}
  \nu(t)
  &\simeq 
  \tilde{\nu}
  +
  \frac{\D \nu}{\D t}
  \bigg|_{t=\tilde{t}}
  \left(
    t
    -
    \tilde{t}
  \right)
  =
  \tilde{\nu}
  \left[
    1
    +
    \tilde{\nu}
    \left(
      t
      -
      \tilde{t}
    \right)
  \right]
  \simeq
  \tilde{\nu}
  \E^{\tilde{\nu}(t-\tilde{t})},
\end{split}
\end{equation}
where $\tilde{\nu}=\nu(\tilde{t})$. 
Then, $R(t)=\exp[-\int_{0}^{t} \D t^{\prime} \nu(t^{\prime})]$ can be approximated to 
\begin{equation}
  R(t) 
  \simeq 
  \exp
  \left\{
    -\Lambda
    \left[
      \exp
      \left(
        \tilde{\nu}t
      \right)
      -
      1
    \right]
  \right\}, 
\end{equation}
where $\Lambda=\E^{-\tilde{\nu}\tilde{t}}$. 
Thus, $\langle I \rangle$ is given by 
\begin{equation}
\begin{split}
  \langle I \rangle
  &\simeq 
  \varkappa
  \E^{\Lambda}
  \int_{0}^{\infty} \D t 
  \exp
  \left(
    -\Lambda
    \E^{\tilde{\nu}t}
  \right)
  =
  \frac{\varkappa \E^{\Lambda}}{\tilde{\nu}}
  E_{1}(\Lambda),
\end{split}
\end{equation}
where $E_{\beta}(z)=\int_{1}^{\infty} \D t \E^{-zt}/t^{\beta}$ is the exponential integral. 
It should be noted that $E_{1}(\Lambda)$ is expanded as [21]
\begin{equation}
  E_{1}(\Lambda)
  =
  -\gamma
  -
  \log \Lambda
  -
  \sum_{k=1}^{\infty}
  \frac{(-\Lambda)^{k}}{kk!},
\end{equation}
where $\gamma=0.57721...$ is the Euler constant. 
In general, the moment $\langle I^{n} \rangle = \int_{0}^{1} \D R I^{n}=n \varkappa^{n}\int_{0}^{\infty}\D t Rt^{n-1}$ is given by 
\begin{equation}
\begin{split}
  \langle I^{n} \rangle 
  &=
  n \varkappa^{n}
  \E^{\Lambda}
  \int_{0}^{\infty} \D t\ 
  t^{n-1}
  \exp
  \left(
    -\Lambda
    \E^{\tilde{\nu}t}
  \right)
\\
  &=
  n 
  \left(
    \frac{\varkappa}{\tilde{\nu}}
  \right)^{n}
  \E^{\Lambda}
  \int_{1}^{\infty} \D x 
  \frac{(\log x)^{n-1}}{x}
  \E^{-\Lambda x}.
\end{split}
\end{equation}
Then, the standard deviation of the current, 
$\sigma_{I} \!=\! \sqrt{\langle I^{2} \rangle \!-\! \langle I \rangle^{2}}$, is given by 
\begin{equation}
\begin{split}
  \sigma_{I}^{2}
  \!=\!
  \left(
    \frac{\varkappa}{\tilde{\nu}}
  \right)^{2}\!
  &
  \left\{
    \frac{\pi^{2}}{6}
    \!+\!
    \Lambda
    \!
    \left[
      \frac{\pi^{2}}{6}
      \!-\!
      2
      \!+\!
      \gamma
      \left(
        2
        \!-\!
        \gamma
      \right)
      \!+\!
      \log \! \Lambda
      \!
      \left(
        2
        \!-\!
        2 \gamma
        \!-\!
        \log \! \Lambda
      \right)
    \right]
  \right.
\\
  &+\!
    \frac{\Lambda^{2}}{2}
    \!
    \left[
      \frac{\pi^{2}}{6}
      \!-\!
      \frac{11}{2}
      \!+\!
      \gamma
      \left(
        7 \!-\! 3\gamma
      \right)
      \!+\!
      \log \! \Lambda
      \!
      \left(
        7 \!-\! 6 \gamma \!-\! 3 \log \! \Lambda
      \right)
    \right]
\\
  &+\!
    \frac{\Lambda^{3}}{3!}
    \!
    \left[
      \frac{\pi^{2}}{6}
      \!-\!
      \frac{247}{18}
      \!+\!
      \frac{7 \gamma (8 \!-\! 3 \gamma) \!+\! 7 \log \! \Lambda \! (8 \!-\! 6 \gamma \!- \! 3 \log \! \Lambda)}{3}
    \right]
\\
  &+
  \left.
    \mathscr{O}(\Lambda^{4})
  \right\}.
\end{split}
\end{equation}

% =========================================================================================== %

Since the thermal stability can be estimated by evaluating the parameter $\tilde{\nu}$, as shown below, 
let us derive the relations between $\tilde{\nu}$ and 
experimentally measurable variables. 
The difference between the most likely switching current $I(\tilde{t})=\varkappa \tilde{t}$ and 
mean switching current $\langle I \rangle$ is given by 
\begin{equation}
  \langle I \rangle 
  -
  I(\tilde{t}) 
  =
  -\frac{\varkappa \E^{\Lambda}}{\tilde{\nu}}
  \left[
    \gamma 
    +
    \sum_{k=1}^{\infty}
    \frac{(-\Lambda)^{k}}{kk!}
  \right]
  +
  \left(
    \E^{\Lambda}
    -
    1
  \right)
  \varkappa
  \tilde{t}.
\end{equation}
For $b=1$, $\tilde{\nu}$ and $\tilde{\nu}\tilde{t}$ are, respectively, given by 
\begin{equation}
  \tilde{\nu}
  =
  \frac{\varkappa \Delta_{0}}{I_{\rm c}},
  \label{eq:nu_b_1}
\end{equation}
\begin{equation}
  \tilde{\nu}
  \tilde{t}
  =
  \Delta_{0}
  \left[
    1
    -
    \frac{1}{\Delta_{0}}
    \log 
    \left(
      \frac{f_{0}I_{\rm c}}{\varkappa \Delta_{0}}
    \right)
  \right].
\end{equation}
As shown in Refs. [22],[23]
$I(\tilde{t})/I_{\rm c}=\tilde{\nu}\tilde{t}/\Delta_{0}$ is around $0.4 \sim 1.0$ 
in the experimentally reasonable temperature and sweep rate regions 
(so called fast pulling regime or Garg's limit [24],[25]). 
Thus, we can approximate that 
$\Lambda =\E^{-\tilde{\nu}\tilde{t}} \simeq 0$ and 
$\E^{\Lambda}=\E^{\E^{-\tilde{\nu}\tilde{t}}}\simeq \E^{\E^{-\Delta_{0}}}\simeq 1$ 
for $\Delta_{0}\gg 1$. 
Then, $\langle I \rangle - I(\tilde{t})$ for $b=1$ is given by 
\begin{equation}
  \langle I \rangle 
  - 
  I(\tilde{t})
  = 
  -\gamma 
  \frac{I_{\rm c}}{\Delta_{0}}.
  \label{eq:diff_b_1}
\end{equation}
Similarly, for $b>1$, 
by using the approximation $\Plog(z) \simeq \log z$, 
$\tilde{\nu}$ and $\tilde{\nu}\tilde{t}$ are, respectively, given by 
\begin{equation}
  \tilde{\nu}
  \simeq
  \left(
    \frac{b-1}{b}
  \right)^{1-1/b}
  \frac{b \varkappa \Delta_{0}^{1/b}}{I_{\rm c}},
  \label{eq:nu_b}
\end{equation}
\begin{equation}
\begin{split}
  \tilde{\nu}
  \tilde{t}
  &\simeq
  \left(
    \frac{b-1}{b}
  \right)^{1-1/b}
  \Delta_{0}^{1/b}
\\
  &\ \ \ \ \ \ \times
  \left(
    1
    -
    \left\{
      \frac{1}{\Delta_{0}}
      \log
      \left[
        \left(
          \frac{b}{b-1}
        \right)^{1-1/b}
        \frac{f_{0}I_{\rm c}}{b \varkappa \Delta_{0}^{1/b}}
      \right]
    \right\}^{1/b}
  \right),
\\
  &\simeq
  \left(
    \frac{b-1}{b}
  \right)^{1-1/b}
  \Delta_{0}^{1/b}
\end{split}
\end{equation}
Then, $\langle I \rangle - I(\tilde{t})$ for $b>1$ is given by 
\begin{equation}
  \langle I \rangle 
  - 
  I(\tilde{t})
  \simeq 
  -\gamma
  \left(
    \frac{b}{b-1}
  \right)^{1-1/b}
  \frac{I_{\rm c}}{b \Delta_{0}^{1/b}}.
  \label{eq:diff_b}
\end{equation}
$[\langle I \rangle - I(\tilde{t})]/I_{\rm c}$ is approximately zero 
for a sufficiently high thermal stability ($\Delta_{0} \gg 1$)
which means a narrow width of the probability density. 
%(the width of the probability density $\sigma_{p}$, defined by 
%$1/\sigma_{p}^{2}=-(\D^{2}\log p/\D I^{2})_{t=\tilde{t}}=[(\D^{2}\nu/\D I^{2})/\nu-2(\nu/\varkappa)^{2}]_{t=\tilde{t}}$, 
%is same order of $\langle I \rangle-I(\tilde{t}$)). 
We also find 
\begin{equation}
  \frac{\tilde{\nu} \left[ \langle I \rangle - I(\tilde{t}) \right]}{\varkappa}
  \simeq
  -\gamma
  =
  -0.57721...
  \label{eq:universal_relation_1}
\end{equation}
\begin{equation}
  \frac{\tilde{\nu} \sqrt{\langle I^{2} \rangle-\langle I \rangle^{2}}}{\varkappa}
  \simeq
  \sqrt{\frac{\pi^{2}}{6}}
  =
  1.28254...
  \label{eq:universal_relation_2}
\end{equation}
for arbitrary $b$ and $\Delta_{0} \gg 1$. 
We numerically verify Eqs. (\ref{eq:universal_relation_1}) and (\ref{eq:universal_relation_2}) 
among the temperature region $0< T \le 500$ K, 
where the values of the parameters are same with those in Fig. \ref{fig:fig1} 
($\Delta_{0}\propto 1/T$ is taken to be 60 for $T=300$ K). 
Equation (\ref{eq:universal_relation_1}) or (\ref{eq:universal_relation_2}) can be used 
to determine the value of $\tilde{\nu}$ experimentally. 
Otherwise, $\tilde{\nu}$ can be estimated by using the relation 
\begin{equation}
  \tilde{\nu}
  =
  -\frac{1}{R}
  \frac{\D R}{\D t}
  \bigg|_{t=\tilde{t}}
  =
  -\frac{\D}{\D t}
  \log R
  \bigg|_{t=\tilde{t}}.
  \label{eq:nu_experiment}
\end{equation}

% =========================================================================================== %

Let us discuss the effect of the value of $b$ on the estimation of the retention time of MRAM. 
We assume that the value of $I_{\rm c}$ is experimentally determined by some other experiments [5].
Then, the unknown parameter in Eq. (\ref{eq:nu_b_1}) or (\ref{eq:nu_b}) is only the thermal stability. 
As mentioned above, 
$\tilde{\nu}$ can be experimentally determined by using 
Eq. (\ref{eq:universal_relation_1}), (\ref{eq:universal_relation_2}), or (\ref{eq:nu_experiment}). 
By setting $\tilde{\nu}(b=1)=\tilde{\nu}(b=2)$, 
we found that the estimated values of the thermal stability with $b=1$ ($\Delta_{1}$) and $b=2$ ($\Delta_{2}$) 
satisfy the relation $\Delta_{1}=\sqrt{2\Delta_{2}}$. 
Let us define the retention time of MRAM by $t^{*}=\E^{\Delta_{0}}/f_{0}$. 
Then, the ratio of the estimated values of the retention time by $b=1$ ($t_{1}^{*}$) and $b=2$ ($t_{2}^{*}$) is 
given by $t_{2}^{*}/t_{1}^{*}=\E^{\Delta_{2}-\sqrt{2\Delta_{2}}}$, 
which is on the order of $10^{21}$ for $\Delta_{2}=60$ 
and increases with increasing $\Delta_{2}$. 
Thus, the determination of the value of $b$ is important 
for the accurate estimation of the retention time of MRAM. 

% =========================================================================================== %

% =========================================================================================== %

\section{Comparison with Theory of Koch \etal}
\label{sec:Comparison with Theory of Koch etal}

In this section, 
we investigate the difference of the value of $b$ 
between Koch \etal [9] and Refs. [10],[11],[12] 
by comparing the solutions of the Fokker-Planck equation, 
and show that $b$ should be two. 
For simplicity, in this section, the current magnitude is assumed to be constant in time [9],[10],[11],[12]. 

% =========================================================================================== %

First of all, it should be mentioned that 
the analytical solution of the switching probability can be obtained 
only for the two special cases. 
The first one is the uniaxially anisotropic system [10]. 
The second one is the in-plane magnetized thin film 
in which the switching path in the thermally activated region is completely limited to the film plane, 
and thus, the effect of the demagnetization field normal to film plane is neglected [11]. 
In these systems, the magnetization dynamics can be described by one variable (the angle from the easy axis, $\theta$), 
although, in general, the magnetization dynamics is described by two angles (the zenith angle $\theta$ and azimuth angle $\varphi$). 
Then, the thermal switching of the magnetization can be regarded as 
the one dimensional Brownian motion of a point particle. 
Although the effect of the demagnetization field of 
an in-plane magnetized system is taken into account 
in the definition of the critical current of Ref. [9], 
the model of Ref. [9] should be regarded as the identical with the models in Refs. [10],[11] 
because the assumption $\mathbf{H}\parallel \mathbf{p}$ in Ref. [9] is valid 
for the two special cases mentioned above, 
where $\mathbf{H}$ and $\mathbf{p}$ are the total magnetic field acting on the free layer 
and magnetization direction of the pinned layer, respectively. 

% =========================================================================================== %

The difficulty to calculate the spin torque assisted thermal switching probability %in general system 
arises from the fact that
the spin torque cannot be expressed as the torque due to the conserved energy. 
Mathematically, it means that we cannot find any function $\tilde{F}(\theta,\varphi)$ 
whose two gradients, $\partial \tilde{F}/\partial \varphi$ and $\partial \tilde{F}/\partial \theta$, 
simultaneously give the spin torque terms of the Landau-Lifshitz-Gilbert equation 
in $(\theta,\varphi)$ coordinate. 
Then, the steady state solution of the Fokker-Planck equation deviates from the Boltzmann distribution. 
However, in the two special cases mentioned above, 
since the magnetization dynamics depends on only $\theta$, 
$\tilde{F}$ can be obtained by integrating the spin torque term with respect to $\theta$. 
Then, the Fokker-Planck equation, 
\begin{equation}
\begin{split}
  \frac{\partial W}{\partial t}
  \!=\!
  \frac{\alpha\gamma^{\prime}}{\sin\theta}
  \frac{\partial}{\partial\theta}
  &
  \!
  \left\{
    \sin\theta
    \!
    \left[\!
      \left(\!\!
        H_{\rm appl}
        \!+\!
        \frac{H_{\rm s}}{\alpha}
        \!+\!
        H_{\rm K}\!
        \cos\theta\!\!
      \right)\!
      \sin\theta
      W
    \right.
  \right.
\\
  &\ \ \ \ \ +\!
  \left.
    \left.
      \frac{k_{\rm B}T}{MV}
      \frac{\partial W}{\partial\theta}
    \right]\!
  \right\},
  \label{eq:FP}
\end{split}
\end{equation}
has a steady state solution of the Boltzmann distribution form, $W \propto \exp[-\mathscr{F}/(k_{\rm B}T)]$.
Here $M$, $V$, $H_{\rm appl}$, $H_{\rm K}$, $H_{\rm s}(\propto I)$, $\gamma_{0}=(1+\alpha^{2})\gamma^{\prime}$, and $\alpha$ are 
the magnetization, volume of the free layer, applied field, uniaxial anisotropy field, 
strength of the spin torque in the unit of the magnetic field, 
gyromagnetic ratio, and the Gilbert damping constant, respectively. 
$F=-MH_{\rm appl}V \cos\theta-(MH_{\rm K}V/2) \cos^{2}\theta$ is the magnetic energy, 
and $\mathscr{F}$ is the effective magnetic energy given by 
\begin{equation}
  \frac{\mathscr{F}}{MV}
  =
  -H_{\rm appl}
  \cos\theta
  -
  \frac{H_{\rm s}}{\alpha}
  \cos\theta
  -
  \frac{1}{2}
  H_{\rm K}
  \cos^{2}\theta.
  \label{eq:free_energy}
\end{equation}
The term $-(MH_{\rm s}V/\alpha)\cos\theta$ in Eq. (\ref{eq:free_energy}) corresponds to $\tilde{F}$ mentioned above. 
By using the steady state solution of the Fokker-Planck equation, 
we can calculate the switching probability, 
according to Refs. [8],[10],[11]. 

% =========================================================================================== %

Koch \etal argued that Brown's formula with the magnetic energy $F$ is applicable 
to the spin torque switching problem 
by replacing $\alpha$ and $T$ 
with $\tilde{\alpha} \!=\! \alpha[1 \!+\! H_{\rm s}/(\alpha H)]$ 
and $\tilde{T} \!=\! T/[1 \!+\! H_{\rm s}/(\alpha H)]$, 
where $H=|\mathbf{H}|=|H_{\rm appl}+H_{\rm K}\cos\theta|$. 
These replacements arise from the assumption that 
the directions of the spin torque ($\propto\! \mathbf{M} \!\times\! (\mathbf{M} \!\times\! \mathbf{p})$) 
and the Landau-Lifshitz damping ($\propto\! \mathbf{M} \!\times\! (\mathbf{M} \!\times\! \mathbf{H})$) are parallel, 
i.e., $\mathbf{H} \parallel \mathbf{p}$. 
At the minimum of the magnetic energy $F$,
$\tilde{T} \!=\! T/(1-I/I_{\rm c})$, 
and thus, Ref. [9] argued that the exponent of the current term of the potential barrier height 
($\propto\! MH_{\rm K}V/(2k_{\rm B}\tilde{T})$) is unity. 
However, it should be noted that 
the definition of the the potential barrier height requires 
not only the minimum of the magnetic energy $F_{\rm min} \!=\! F(0)$ 
but also its maximum $F_{\rm max} \!=\! F(\theta_{\rm m})$ divided by the temperature, 
where $\theta_{\rm m}=\cos^{-1}(-H_{\rm appl}/H_{\rm K})$. 
We can easily verify that $H$, and also $\tilde{T}$, are zero at $\theta=\theta_{\rm m}$. 
Thus, $F_{\rm max}/[k_{\rm B}\tilde{T}(\theta_{\rm m})]$ is not well defined, 
and the relation argued in Ref. [9] is not satisfied, as shown below:
\begin{equation}
  \frac{F_{\rm max}}{k_{\rm B}\tilde{T}(\theta_{\rm m})}
  -
  \frac{F_{\rm min}}{k_{\rm B}\tilde{T}(0)}
  \neq 
  \frac{(F_{\rm max}-F_{\rm min})(1-I/I_{\rm c})}{k_{\rm B}T}.
\end{equation}

% =========================================================================================== %

The origin of the problem in Ref. [9] is that 
$\exp[-F/(k_{\rm B}\tilde{T})]$ is not a steady state solution 
of the Fokker-Planck equation (\ref{eq:FP}):
the steady state solution is $\exp[-\mathscr{F}/(k_{\rm B}T)]$. 
Since the effect of the spin torque can be regarded as an additional term to the applied field, 
as shown in Eq. (\ref{eq:free_energy}), 
the potential barrier height of the spin torque assisted thermal switching is, 
similar to Brown's formula [8], given by 
\begin{equation}
  \frac{\mathscr{F}_{\rm max}-\mathscr{F}_{\rm min}}{k_{\rm B}T}
  =
  \Delta_{0}
  \left(
    1
    +
    \frac{H_{\rm appl}+H_{\rm s}/\alpha}{H_{\rm K}}
  \right)^{2},
\end{equation}
where the thermal stability is defined by $\Delta_{0}=MH_{\rm K}V/(2k_{\rm B}T)$. 
By using the relation 
\begin{equation}
  \left(
    1
    \!+\!
    \frac{H_{\rm appl} \!+\! H_{\rm s}/\alpha}{H_{\rm K}}
  \right)
  \!=\!
  \left(
    1
    \!+\!
    \frac{H_{\rm appl}}{H_{\rm K}}
  \right)
  \!\!
  \left[
    1
    \!+\!
    \frac{H_{\rm s}}{\alpha(H_{\rm K} \!+\! H_{\rm appl})}
  \right],
\end{equation}
and defining the critical current $I_{\rm c}$ by $H_{\rm s}/[\alpha(H_{\rm K}+H_{\rm appl})]=-I/I_{\rm c}$, 
we find that [11]
\begin{equation}
  \frac{\mathscr{F}_{\rm max}-\mathscr{F}_{\rm min}}{k_{\rm B}T}
  =
  \Delta_{0}
  \left(
    1
    +
    \frac{H_{\rm appl}}{H_{\rm K}}
  \right)^{2}
  \left(
    1
    -
    \frac{I}{I_{\rm c}}
  \right)^{2},
\end{equation}
Thus, the exponent of the current term should be two. 

% =========================================================================================== %

% =========================================================================================== %

\section{Conclusion}
\label{sec:Conclusion}

In conclusion, 
we studied the spin torque assisted thermal switching of the single free layer theoretically. 
We derived the theoretical formulas of 
the most likely and averaged switching currents 
of the sweep current assisted magnetization reversal, 
and showed that the value of the exponent $b$ in the switching rate 
significantly affects the estimation of the retention time of MRAM. 
We also discussed the difference between the theories in Ref. [9] and Refs. [10],[11] 
from the Fokker-Planck approach, 
and showed that the exponent should be two. 

% =========================================================================================== %

\section*{Acknowledgment}

The authors would like to acknowledge 
H. Kubota and S. Yuasa 
for the valuable discussions they had with us. 

% =========================================================================================== %

\ifCLASSOPTIONcaptionsoff
  \newpage
\fi

% =========================================================================================== %

\end{document}